\documentclass[aps,pra,reprint,groupedaddress]{revtex4-1}

\usepackage{graphicx}
\usepackage{multirow}
\usepackage{mathrsfs}
\usepackage{amsfonts}
\usepackage{amssymb}
\usepackage{cancel}
\usepackage{amsmath}
\usepackage{bm}
\usepackage{latexsym}
\usepackage{color}

\begin{document}


\title{An inverse-system method for identification of damping rate functions in non-Markovian quantum systems}


\author{Shibei Xue$^{1,2}$, Lingyu Tan$^{1,2}$, Rebing Wu$^{3,4}$, Min Jiang$^{5}$, Ian~R.~Petersen$^{6}$}
\affiliation{$^1$Department of Automation, Shanghai Jiao Tong University, Shanghai 200240, P. R. China\\
$^2$Key Laboratory of System Control and Information Processing, Ministry of Education of China, Shanghai 200240, P. R. China\\
$^3$Department of Automation, Tsinghua University, Beijing 100084, P. R. China\\
$^4$Center for Quantum Information Science and Technology, TNList, Beijing 100084,
P. R. China\\
$^5$School of Electronic and Information Engineering, Soochow University, Suzhou 215006 , P. R. China\\
$^6$Research School of Electrical, Energy and Materials Engineering, The Australian National University, Canberra 0200,
Australia}\email[]{shbxue@sjtu.edu.cn}


\date{\today}

\begin{abstract}
Identification of complicated quantum environments lies in the core of quantum engineering, which systematically constructs an environment model with the aim of accurate control of quantum systems. In this paper, we present an inverse-system method to identify damping rate functions which describe non-Markovian environments in time-convolution-less master equations. To access  information on the environment, we couple a finite-level quantum system to the environment and measure time traces of local observables of the system. By using sufficient measurement results, an algorithm is designed, which can simultaneously estimate multiple damping rate functions for different dissipative channels. Further, we show that identifiability for the damping rate functions corresponds to the invertibility of the system and a necessary condition for identifiability is also given. The effectiveness of our method is shown in examples of an atom and three-spin-chain non-Markovian systems.
\end{abstract}

\pacs{}

\maketitle

\section{Introduction}
In the past two decades, significant progress has been achieved in quantum computing~\cite{steane1998quantum}, quantum communication~\cite{gisin2007quantum}, and quantum metrology~\cite{giovannetti2006quantum}, etc., which rely on precise manipulation of quantum systems.
To this end, many control methods have been proposed such as optimal control~\cite{doi:10.1021/j100149a003}, feedback control~\cite{PhysRevA.62.022108, xue2017non}, linear quadratic Gaussian control~\cite{NURDIN20091837, xue2017feedback}, which are designed based on the exact models of quantum systems. However, these models may not be well constructed due to unknown dynamics. For example, in an experiment on a quantum dot system, a calculated curve based on a master equation model has a large discrepency from the experimental data~\cite{PhysRevLett.108.046807}. Hence, systematic methods for calibration of a model of a quantum system are required.

These methods refer to system identification, which utilizes the inputs and outputs of a quantum system to specify an exact model of a quantum system. Quantum system identification was firstly developed for closed quantum systems, including molecular systems~\cite{phan1997learning,geremia2002optimal}, two-level systems~\cite{schirmer2004experimental,bonnabel2009observer}, spin networks~\cite{d2005controllability,albertini2005model,burgarth2009indirect,burgarth2012quantum,zhang2014quantum,hou2017experimental,wang2018quantum}.
However, practical systems are open to quantum environments. So identification methods were developed for Markovian quantum systems which are in a memoryless environment. In~\cite{mabuchi1996dynamical}, a continuous-measurement approach was proposed to access the information of a cavity QED system and thus unknown parameters in the Hamiltonian can be identified in a maximum-likelihood sense. This method was generalized to the identification of parameters in an atom~\cite{gambetta2001state} or a two-level system~\cite{xue2013identification} and of even unknown structures in spin networks~\cite{kato2014structure}.
Also, in the problem of Hamiltonian identification, a class of similarity-realization methods was presented for the quantum system with a linear algebraic structure. In~\cite{zhang2015identification}, measured time traces of local observables of Markovian spin networks were utilized to construct a realization of the system in the coherence vector representation so as to identify unknown parameters by solving a set of nonlinear equations which are established according to an equivalence between the realization and the original system. Its identifiability was discussed in~\cite{sone2017hamiltonian}.
Similarly, for a linear quantum system, its realization can be obtained from the spectrum of a probing field and thus unknown parameters can be identified when there exists a similarity map to transform the realization to a physical-realizable quantum system~\cite{guctua2016system,levitt2017identification}.

However, the above existing methods were challenged when the environment of a quantum system exhibits memory effects,
because this kind of environment results in complicated non-Markovian dynamics of quantum systems~\cite{breuer2016colloquium}.
For these quantum systems, a frequency-domain-analysis method was proposed to identify the noise spectrum for a non-Markovian superconducting qubit in a special state~\cite{wu2013spectral}. Also, an augmented system model was presented to estimate the noise spectrum of a quantum dot system where linear ancillary systems were introduced to represent the internal modes of the environment~\cite{8820133}. A similar problem was considered for non-Markovian quantum systems described by a class of time-convolution-less (TCL) master equations where the environment is characterized by a time-varying damping rate function. Hence, identification of the environment is converted to reconstructing the damping rate function. For a non-Markovian single qubit system, a least square identifier was designed where the function can be expanded in a time series and thus the corresponding coefficients can be estimated in a least square sense~\cite{Xue19QIP}. In addition, a gradient algorithm was designed to identify the function for spin chains, whose computational cost is heavy due to calculation of the system dynamics for multiple times~\cite{xue2019identification}. Hence, effective algorithms for solving the identification of the damping rate functions in general non-Markovian quantum systems should be designed.
%

In this paper, we present an inverse-system method to identify the damping rate function in the TCL master equation for describing the non-Markovian environment. We measure the time traces of observables of the non-Markovian system whose derivative can be expressed in terms of the damping rate function. When this relation is invertible, the damping rate function can be represented by the time traces in a least square sense. We also discuss the identifiability for this problem from a perspective of inverse systems where a necessary condition for identifiability is obtained. Based on the analysis on identifiability, a numerical algorithm is designed, whose effectiveness is verified in two physical examples.

This paper is organized as follows. In Section~\ref{III} we introduce a time-convolution-less master equation for describing the dynamics of non-Markovian quantum systems. We present our inverse-system method for the identification of damping rate functions in Section \ref{idenProcess} where identifiability and design of a numerical algorithm are also discussed. In addition, in Section \ref{examples}, we give two examples to show the effectiveness of our method. Finally, we draw our conclusions in Section \ref{VI}.

\section{Non-Markovian quantum system model}\label{III}
The dynamics of a quantum system interacting with non-Markovian environments can be described by a TCL master equation
\begin{equation}\label{7}
 {\dot{\rho}}(t)=\mathcal{L}_0{\rho}(t)+\vec{\gamma}^T(t){\vec{\mathcal{L}}_{\gamma}}{\rho}(t),
\end{equation}
where ${\rho}(t)$ is the density matrix of the quantum system.
The first term on the RHS of Eq.~(\ref{7})
\begin{equation}\label{7-1}
  \mathcal{L}_0{\rho}(t)=-i[{H},{\rho}(t)]
\end{equation}
describes the internal dynamics of the quantum system whose  Hamiltonian is ${H}$. Also, the second term describes the dissipative processes induced by $N$ independent non-Markovian environments, where
\begin{equation}\label{7-2}
{\vec{\mathcal{L}}_{\gamma}}{\rho}(t)=[\mathcal{L}_{1}{\rho}(t),\cdots,\mathcal{L}_{n}{\rho}(t),\cdots,\mathcal{L}_{N}{\rho}(t)]^T
\end{equation}
 with each dissipative channel in a Lindblad form
\begin{equation}\label{8}
  \mathcal{L}_{n}{\rho}(t)={L}_n{\rho}(t){L}_n^\dagger-\frac{1}{2}{L}_n^\dagger {L}_n{\rho}(t)-\frac{1}{2}{\rho}(t){L}_n^\dagger {L}_n
\end{equation} and the coupling operator ${L}_n$.

The non-Markovian properties of the environments, such as the density states and the noise spectra of the environments, are embedded in a vector of the damping rate functions
\begin{equation}\label{14}
 \vec{\gamma}(t)=[\gamma_1(t), \cdots,\gamma_n(t),\cdots,\gamma_N(t)]^T,
\end{equation}
where $\gamma_n(t)$ is a time-varying damping rate function for the $n$th dissipative channel.

\section{An inverse-system method for identification of the damping rate functions}\label{idenProcess}
\subsection{Problem Formulation}
The damping rate functions in the TCL master equation contain all of the information about the non-Markovian environment including the power spectral density, the coupling strengths and the density of states of the environment, which determine the exact dynamics of non-Markovian quantum systems. As shown in \cite{breuer2002theory}, an atom system described by a TCL master equation with a damping rate function exhibits an oscillating dissipative process which is quite different from its corresponding Markovian dynamics. However, in some circumstances, the damping rate functions may not be known, for example, in a quantum dot coupled to an unknown quantum colored noise in~\cite{PhysRevLett.108.046807}.

Generally, the environment of quantum systems cannot be directly measured since it is complicated and not directly accessible. Alternatively, we assume that we have many copies of the quantum system such that an observable of the quantum system can be measured many times so as to obtain the time trace of the observable. In this way, we may evaluate the environment indirectly. In this paper, we measure $M$ observables of the non-Markovian quantum system
\begin{equation}\label{14-1}
  \vec{O}=[\begin{array}{ccccc}
            O_1 &\cdots& O_m &\cdots& O_M
          \end{array}]^T.
\end{equation}
Here, we consider the measurement results as the outputs of the system, which can be expressed as
\begin{equation}\label{9}
\vec{y}(t)=\left[
       \begin{array}{cccccc}
         y_1(t), &  \cdots& y_m(t), &\cdots,& y_M(t)
       \end{array}
     \right]^T
\end{equation}
with
\begin{equation}\label{9-1}
  y_m(t)={\rm tr}[{O}_m{\rho}(t)].
\end{equation}

With these measurements, we consider the following problem for accurate identification of the damping rate functions in the TCL master equation.

Given a non-Markovian quantum system described by the TCL master equation (\ref{7}), we utilize measurements of the time traces of the observables  (\ref{9}) to identify $N$ unknown damping rate functions $\{\gamma_n(t), n=1,2,\cdots,N\}$.
\subsection{An inverse-system method for the identification problem}
In this paper, we present an inverse-system approach to identifying the damping rate functions. In the non-Markovian quantum system, the unknown damping rate functions result in the time trace of the observables (\ref{9}) and thus the measurements (\ref{9}) provide information about $\vec{\gamma}(t)$; i.e., we can consider the measurement result as a function of  the damping rate function $\vec{y}(t)=f(\vec{\gamma}(t))$.
The non-Markovian quantum system (\ref{7}) is said to be left invertible if there exists a function $g$ such that $g\circ f(\vec{\gamma}(t))=\vec{\gamma}(t)$; i.e., the composite function $g\circ f$ is the identity function. Here, $g$ is the inverse function of $f$.

In our method, we try to estimate $\vec{\gamma}(t)$ by seeking the function $g$. In other words, the damping rate function $\vec{\gamma}(t)$ can be identified if we can express $\vec{\gamma}(t)$ as a function of the measurement results; i.e., $\vec{\gamma}(t)=g(\vec{y}(t))$. We will follow this basic idea to design our identification method and this is the reason that we call it an inverse-system method.


We should firstly express the vector of damping rate functions $\vec{\gamma}(t)$ in terms of the measurements (\ref{9}). Using the equation (\ref{7}), we differentiate the output (\ref{9-1}) and thus obtain
\begin{eqnarray}\label{10}
\dot y_1(t)&=&  \langle \mathcal{L}^{*}_0{O}_1\rangle+\gamma_1(t)\langle \mathcal{L}^{*}_{1}{O}_1\rangle+\gamma_2(t)\langle \mathcal{L}^{*}_{2}{O}_1\rangle+\cdots+\nonumber\\
&&\gamma_N(t)\langle \mathcal{L}^{*}_{N}{O}_1\rangle\nonumber\\
\dot y_2(t)&=&  \langle \mathcal{L}^{*}_0{O}_2\rangle+\gamma_1(t)\langle \mathcal{L}^{*}_{1}{O}_2\rangle+\gamma_2(t)\langle \mathcal{L}^{*}_{2}{O}_2\rangle+\cdots+\nonumber\\
&&\gamma_N(t)\langle \mathcal{L}^{*}_{N}{O}_2\rangle\nonumber\\
&\vdots&\nonumber\\
\dot y_M(t)&=&  \langle \mathcal{L}^{*}_0{O}_M\rangle+\gamma_1(t)\langle \mathcal{L}^{*}_{1}{O}_M\rangle+\gamma_2(t)\langle \mathcal{L}^{*}_{2}{O}_M\rangle+\cdots\nonumber\\
&&+\gamma_N(t)\langle \mathcal{L}^{*}_{N}{O}_M\rangle,
\end{eqnarray}
where we have rewritten the corresponding terms by using the relation ${\rm tr}[{O}_m\mathcal{L}_n{\rho}]={\rm tr}[{\rho}\mathcal{L}^{*}_n{O}_m]=\langle \mathcal{L}^{*}_n{O}_m\rangle$  with
\begin{equation}\label{}
  \mathcal{L}^{*}_n{O}_m=\left\{\begin{array}{cc}
                        -i[{O}_m,{H}], & n=0 \\
                       L_n^\dagger {O}_mL_n-\frac{1}{2}L_n^\dagger L_n{O}_m & \\
                      -\frac{1}{2}{O}_m L_n^\dagger L_n, &n=1,2,\cdots,N.
                      \end{array}\right.
\end{equation}
We then rewrite the equation (\ref{10}) in a vector form as
\begin{equation}\label{11}
  \vec{b}(t)=\bar{W}(t)\vec{\gamma}(t),
\end{equation}
where $\vec{b}(t)\in\mathbb{R}^M$ and $\bar{W}(t)\in\mathbb{R}^{M\times N}$ are expressed as
\begin{equation}\label{12}
    \vec{b}(t)= \left[
          \begin{array}{c}
             \dot y_1(t)- \langle \mathcal{L}^{*}_0{O}_1\rangle\\
            \vdots \\
            \dot y_M(t)- \langle \mathcal{L}^{*}_0{O}_M\rangle\\
          \end{array}
        \right],
\end{equation}
\begin{equation}\label{13}
  \bar{W}(t)=\langle \vec{\mathcal{L}}^{*}_\gamma\vec{O}\rangle=\left[
         \begin{array}{c}
           \langle \vec{\mathcal{L}}^{*}_{\gamma}{O}_1\rangle \\
            \vdots \\
          \langle \vec{\mathcal{L}}^{*}_{\gamma}{O}_M\rangle \\
         \end{array}
       \right]
\end{equation}
with
\begin{equation}\label{13-1}
 \langle\vec{\mathcal{L}}_\gamma^{*}O_m\rangle=[\begin{array}{ccc}
                                             \langle\mathcal{L}^{*}_{1}O_m \rangle, &\cdots, & \langle\mathcal{L}^{*}_{N}O_m \rangle
                                           \end{array}],
\end{equation}
respectively.

The system is left invertible if $\bar{W}(t)$ is of full column rank, i.e. $ rank(\bar{W}(t))=N$. We can express $\vec{\gamma}(t)$ in a least square sense as
\begin{equation}\label{15}
  \vec{{\gamma}}(t)=(\bar{W}^T(t)\bar{W}(t))^{-1}\bar{W}^T(t)\vec{b}(t).
\end{equation}

Further, substituting the expression for $\vec{\gamma}(t)$ (\ref{15}) into the TCL master equation (\ref{7}), we have
\begin{eqnarray}\label{16}
  {\dot{{\rho}}}(t)&=&\mathcal{M}(t){{\rho}}(t)\nonumber\\
  &=&\mathcal{L}_0{{\rho}}(t)+{\Big (}(\bar{W}^T(t)\bar{W}(t))^{-1}\bar{W}^T(t)\vec{b}(t){\Big )}^T \vec{\mathcal{L}}_{\gamma}{{\rho}}(t)\nonumber\\
\end{eqnarray}
where the dynamics of ${{\rho}}(t)$ are determined by the measurements results.
The master equation (\ref{16}) can be formally solved as
\begin{equation}\label{16-1}
  {\rho}(t) = \exp {\Big \{}\int_0^t\mathcal{M}(\tau)d\tau{\Big \}}{\rho}(0)
  \end{equation}
where  ${\rho}(0)$ is the initial
density matrix of the system.

Finally, with the above derivation, we can formally identify $\vec{{\gamma}}(t)$ with the expression (\ref{15}) and (\ref{16}). Note that the expression (\ref{15}) depends on the choice of the observables. We will analyze how to choose these observables in the next subsection.

\subsection{Identifiability analysis for the damping rate functions}
To analyze the identifiability of the damping rate functions, we consider the rank of $\vec{\mathcal{L}}^{*}_\gamma\vec{O}$ rather than that of $\bar{W}(t)$ (\ref{13}) since the time traces of the corresponding observables involve the time-varying density matrix of the system as shown in (\ref{13}). The rank of $\vec{\mathcal{L}}^{*}_\gamma\vec{O}$ is defined as the minimal number of linear independent operator arrays in $\vec{\mathcal{L}}^{*}_\gamma\vec{O}$. This means that the independent operator arrays satisfy
\begin{equation}\label{16-2}
\beta_1\vec{\mathcal{L}}^{*}_{\gamma}{O}_{i_1}+\cdots+\beta_p\vec{\mathcal{L}}^{*}_{\gamma}{O}_{i_p}\neq 0
\end{equation}
for any non-zero real numbers $\beta_1,\cdots,\beta_p$.


When $\vec{\mathcal{L}}^{*}_\gamma\vec{O}$ is full-column-rank, ${\vec \gamma}(t)$ can be identified.
Otherwise, when the rank of $\vec{\mathcal{L}}^{*}_\gamma\vec{O}$ is less than $N$, we can divide $\vec{{O}}$ into $[\begin{array}{cc}
                               \vec{{O}}_1 & \tilde{{O}}_1
                              \end{array}
]^T$ such that
\begin{equation}\label{rankrank}
rank({\vec{\mathcal{L}}^{*}_{\gamma}}\vec{{O}}_1)=rank({\vec{\mathcal{L}}^{*}_{\gamma}}\vec{{O}}).
\end{equation}
Hence, ${\vec{\mathcal{L}}}^{*}_{{\gamma}}\tilde{{O}}_1$ can be expressed in terms of ${\vec{\mathcal{L}}}^{*}_{{\gamma}}\vec{{O}}_1$; i.e,
\begin{equation}\label{o1ot}
{\vec{\mathcal{L}}}^{*}_{{\gamma}}\tilde{{O}}_1=\bar{V}_{11}{\vec{\mathcal{L}}}^{*}_{{\gamma}}\vec{{O}}_1.
\end{equation}

Thus the corresponding outputs can be divided as
\begin{equation}
\vec{y}=\left[\begin{array}{c}\vec{{y}}_1\\
\tilde{{y}}_1\end{array}\right]=\left[\begin{array}{c}\langle\vec{{O}}_1\rangle\\
\langle\tilde{{O}}_1\rangle\end{array}\right],
\end{equation}
whose derivative can be re-expressed as
\begin{eqnarray}
\dot{\vec{{y}}}_1&=&\langle {\mathcal{L}^{*}_0}\vec{{O}}_1\rangle+ \langle {{\vec{\mathcal{L}}}^{*}_{{\gamma}}}\vec{{O}}_1\rangle {\vec\gamma}(t),\label{inde}\\
\dot{\tilde{{y}}}_1&=&\langle {\mathcal{L}^{*}_0}\tilde{{O}}_1\rangle+ \langle {{\vec{\mathcal{L}}}^{*}_{{\gamma}}}\tilde{{O}}_1\rangle{\vec\gamma}(t).\label{noninde}
\end{eqnarray}
Multiplying $\bar{V}_{11}$ on both sides of (\ref{inde}) and subtracting (\ref{inde}) from (\ref{noninde}), we obtain $\dot{\tilde{{y}}}_1-\bar{V}_{11}\dot{\vec{{y}}}_1=\langle{\mathcal{L}^{*}_0}\tilde{{O}}_1-\bar{V}_{11}{\mathcal{L}^{*}_0}\vec{{O}}_1\rangle$,
where we have used the relation (\ref{o1ot}).

Letting $\bar{{y}}_2=\dot{\tilde{{y}}}_1-\bar{V}_{11}\dot{\vec{{y}}}_1$ and $\bar{{O}}_2={\mathcal{L}^{*}_0}\tilde{{O}}_1-\bar{V}_{11}{\mathcal{L}^{*}_0}\vec{{O}}_1$, we obtain $\bar{{y}}_2=\langle\bar{{O}}_2\rangle$. The derivative of $\bar{{y}}_2$ can be written as
\begin{equation}\label{y2}
\dot{\bar{{y}}}_2=\langle{\mathcal{L}^{*}_0}\vec{{O}}_2\rangle+\langle{\vec{\mathcal{L}}^{*}}_{{\gamma}}\vec{{O}}_2\rangle\vec{\gamma}(t).
\end{equation}
If we can find $N$ linear independent operator arrays in
$\left[\begin{array}{cc}
                                                           \vec{\mathcal{L}}^{*}_{{\gamma}}\vec{{O}}_1 & \vec{\mathcal{L}}^{*}_{{\gamma}}\bar{{O}}_2
                                                          \end{array}\right]^T$, the system is left invertible and thus $\vec{\gamma}(t)$ can be recovered from $\vec{{y}}_1$ and $\bar{{y}}_2$.
Otherwise, we can split $\bar{{O}}_2$ into two parts and follow a similar procedure from (\ref{rankrank}) to (\ref{y2}) to span the operator space.             
%
If we can terminate the above iterative procedure within a finite number of steps, the system is invertible. We can obtain a transform $\vec O'=\bar V\vec O=\left[\begin{array}{cccc}
          \vec{{O}}_1  & \vec{{O}}_2 & \cdots & \vec{{O}}_{\alpha}
         \end{array}\right]^T$ such that $rank(\vec O')=N$.
We define $\alpha$ as the relative degree.

Therefore, the system is invertible only when $\alpha$ is a finite number. With the observable $\vec O'$, we have
\begin{equation}
  {\vec{b}}'(t)=\bar{W}'(t)\bar{\gamma}(t)
\end{equation}
where
\begin{equation}
  {\vec{b}}'(t)=\left[\begin{array}{c}
                         f_1(\vec{{y}}^{(1)})-\langle{\mathcal{L}^{*}_0}\vec{{O}}_1\rangle \\
                         f_2(\vec{{y}}^{(1)},\vec{{y}}^{(2)})- \langle{\mathcal{L}^{*}_0}\vec{{O}}_2\rangle\\
                         \vdots \\
                         f_{\alpha}(\vec{{y}}^{(1)},\vec{{y}}^{(2)},\cdots,\vec{{y}}^{(\alpha)})-\langle{\mathcal{L}^{*}_0}\vec{{O}}_{\alpha}\rangle
                       \end{array}\right]
\end{equation}
with $\vec y_k=f_k(\vec{{y}}^{(1)},\vec{{y}}^{(2)},\cdots,\vec{{y}}^{(k)})$, $k=1,2,\cdots,\alpha$,
and
\begin{equation}
  {\bar{W}}'(t)=\left[\begin{array}{c}
                        \langle{\vec{\mathcal{L}}}^{*}_{{\gamma}}\vec{{O}}_1\rangle   \\
                         \langle{\vec{\mathcal{L}}}^{*}_{{\gamma}}\vec{{O}}_2\rangle  \\
                         \vdots  \\
                         \langle{\vec{\mathcal{L}}}^{*}_{{\gamma}}\vec{{O}}_{\alpha}\rangle
                       \end{array}\right].
\end{equation}%
Hence, we can identify  ${\vec{\gamma}}(t)$ as
\begin{equation}\label{esti2}
   {\vec{\gamma}}(t)=({\bar{W'}}^{T}(t)\bar{W}'(t))^{-1}{\bar{W'}}^{T}(t)\bar{b}'(t)
\end{equation}
Note that the above identifiability analysis is based on the rank of $\vec{\mathcal{L}}^{*}_\gamma\vec{O}$ rather than that of $\bar{W}(t)$. Hence, it is a necessary condition for the identifiability we obtained. We aware that a similar analysis was recently given in~\cite{Cao}.
\subsection{A numerical algorithm for identification based on the inverse-system method}
Based on the identifiability analysis in the above subsection, we assume that the initial observables satisfy the necessary condition when we design the following algorithm. 

Due to the presence of an exponential of the integral of
the time-varying superoperator $\mathcal{M}$, it
is difficult to obtain an analytical expression for (\ref{16-1}) in general and thus numerical algorithms are required for the identification based on the inverse-system method. To this end, we discretize a total time $T$ into $K$ segments with a sampling interval of $\Delta=T/K$. This time interval $\Delta$ is also the sampling time in the measurement process, where we assume that we have many copies of the quantum system such that the outputs can be obtained from many measurements on the observables. We denote the measurement data as $\{y_m(k\Delta),m=1,2,\cdots,M,~k=0,1,\cdots,K.\}$

In addition, we assume that $\vec{{\gamma}}(t)$ is a piecewise constant function. The damping rate function as $\vec{{\gamma}}_{k}$ at the $k$th sampling time with $k=0,1,\cdots,K$, can be identified as
\begin{equation}\label{16-1-1}
  \vec{{\gamma}}_{k}=({\bar W}_k^T{\bar W}_k)^{-1}{\bar W}^T_k\vec{b}_k.
\end{equation}
The elements of  ${\bar W}_k$ and $\vec{b}_k$ are expressed as
\begin{eqnarray}
  {{\bar W}}^{mn}_k &=& {\rm tr}[\mathcal{L}^{*}_{n}{O}_m(\cdot)] \\
 {b}^m_k &=& \frac{y_m((k+1)\Delta)-y_m(k\Delta)}{\Delta} -{\rm tr}[\mathcal{L}^{*}_{0}{O}_m(\cdot)]
\end{eqnarray}
respectively, where we have approximated the derivative of the outputs by the first-order forward difference of the outputs. Note that this approximation works well when the sampling time $\Delta$ approaches to zero.

On the other hand, since $\vec{{\gamma}}_{k}$ is constant in each time interval, the superoperator $\mathcal{M}(t)$ is time-invariant such that we can calculate ${\rho}(k\Delta)$ as
\begin{equation}\label{16-2}
 {\rho}(k\Delta)=\hat{\mathcal{M}}_{k-1}\cdots\hat{\mathcal{M}}_1\hat{\mathcal{M}}_0 {\rho}(0),~k=1,\cdots,K,
\end{equation}
where the superoperator $\hat{\mathcal{M}}_\kappa$ is calculated as
\begin{equation}\label{16-3}
\hat{\mathcal{M}}_\kappa=\exp {\Big \{}\Delta{\Big (}\mathcal{L}_0(\cdot)+\vec{{\gamma}}_{\kappa}^T\vec{\mathcal{L}_{\gamma}}(\cdot){\Big )}{\Big \}},\kappa=0,\cdots,k-1.
\end{equation}

Eventually, we can summarize our numerical method based on the inverse-system method as follows.

Step 1: For a given non-Markovian quantum system, initialize the density matrix ${\rho}(0)$, the sampling time $\Delta$ and the final time $T$ and then measure the outputs of the system $\vec{y}(t)$ for a set of given observables $\{{O}_m, m=1,\cdots,M\}$;

Step 2: Identify $\vec{{\gamma}}_{k}$ according to (\ref{16-1-1});

Step 3: Calculate ${\rho}(k\Delta)$ according to (\ref{16-2});

Step 4: Determine whether $T=k\Delta$. If yes, stop the algorithm; otherwise, let $k=k+1$ and go to Step 2.

Note that since the identifiability we obtained is a necessary condition, numerically, there would exist some critical time instants where the full rank condition cannot be satisfied. The identifiability analysis tells us that we should construct new outputs based on the existing measurements to make $\bar{W}_k$ be invertible. However, the procedure to obtain new outputs in the analysis would be indirect. Instead, we can substitute the observable lowering the rank of $\bar{W}_k$ for new corresponding observables such that $\bar{W}_k$ is invertible. It is possible to obtain time traces of new observables since we have assumed that we have many copies of the system. This will be shown in the examples. Moreover, we emphasize that our algorithm only requires us to solve the TCL master equation once as shown in the procedure of our algorithm, which would save computational time.

\section{Two physical examples}\label{examples}
In this section, we will consider two examples for an atom system and a three-spin-chain system, respectively, in non-Markovian environments. The dynamics of both systems are described by TCL master equations, where the damping rate functions are identified from measurements of the observables of the two systems.
%

\subsection{Non-Markovian environment identification for a single-atom system}
In the first example, we consider an example of an atom system where the atom is coupled
to a non-Markovian environment through a single mode cavity~\cite{breuer2002theory}. Here, the atom can be considered as a two-level system whose Hamiltonian is written as ${H}_a=\frac{1}{2}\omega_q{\sigma_z}$ with a splitting frequency $\omega_q=1{\rm GHz}$. The evolution of the atom in a non-Markovian environment can be described by the TCL master equation as
\begin{eqnarray}\label{1}
  {\dot{\rho}}(t)&=&-i[{H}_a,{\rho}(t)]+\gamma_a(t){\Big(}{\sigma}_-{\rho}(t){\sigma}_+-\frac{1}{2}{\sigma}_+{\sigma}_-{\rho}(t)\nonumber\\
  &&-\frac{1}{2}{\rho}(t){\sigma}_+ {\sigma}_-{\Big)}
\end{eqnarray}
with a damping rate function $\gamma_a(t)$. Here, ${\sigma}_+$ and ${\sigma}_-$ are the ladder operators for the atom.

To simulate the real non-Markovian dynamics of the atom, we introduce a damping rate function
\begin{equation}\label{}
\gamma_a(t)=\frac{2\gamma_{0a}\lambda_a\sinh(d_at/2)}{d_a\cosh(d_at/2)+\lambda_a\sinh(d_at/2)}
\end{equation}
with the parameters $\gamma_{0a} = 0.5{\rm GHz}$, $\lambda_a = 0.1{\rm GHz}$, and $d_a = 0.6{\rm GHz}$~\cite{breuer2002theory}. The initial density matrix is set to be $\rho(0)=\frac{1}{2}(I+\frac{1}{\sqrt{3}}\sigma_x+\frac{1}{\sqrt{3}}\sigma_y+\frac{1}{\sqrt{3}}\sigma_z)$.
We sample the observable in a total time $10\mu s$ by ten thousands times. Thus we can calculate the time traces of the observables of the atom. 

%

\begin{figure}
	\centering
	\includegraphics[width=2.8in]{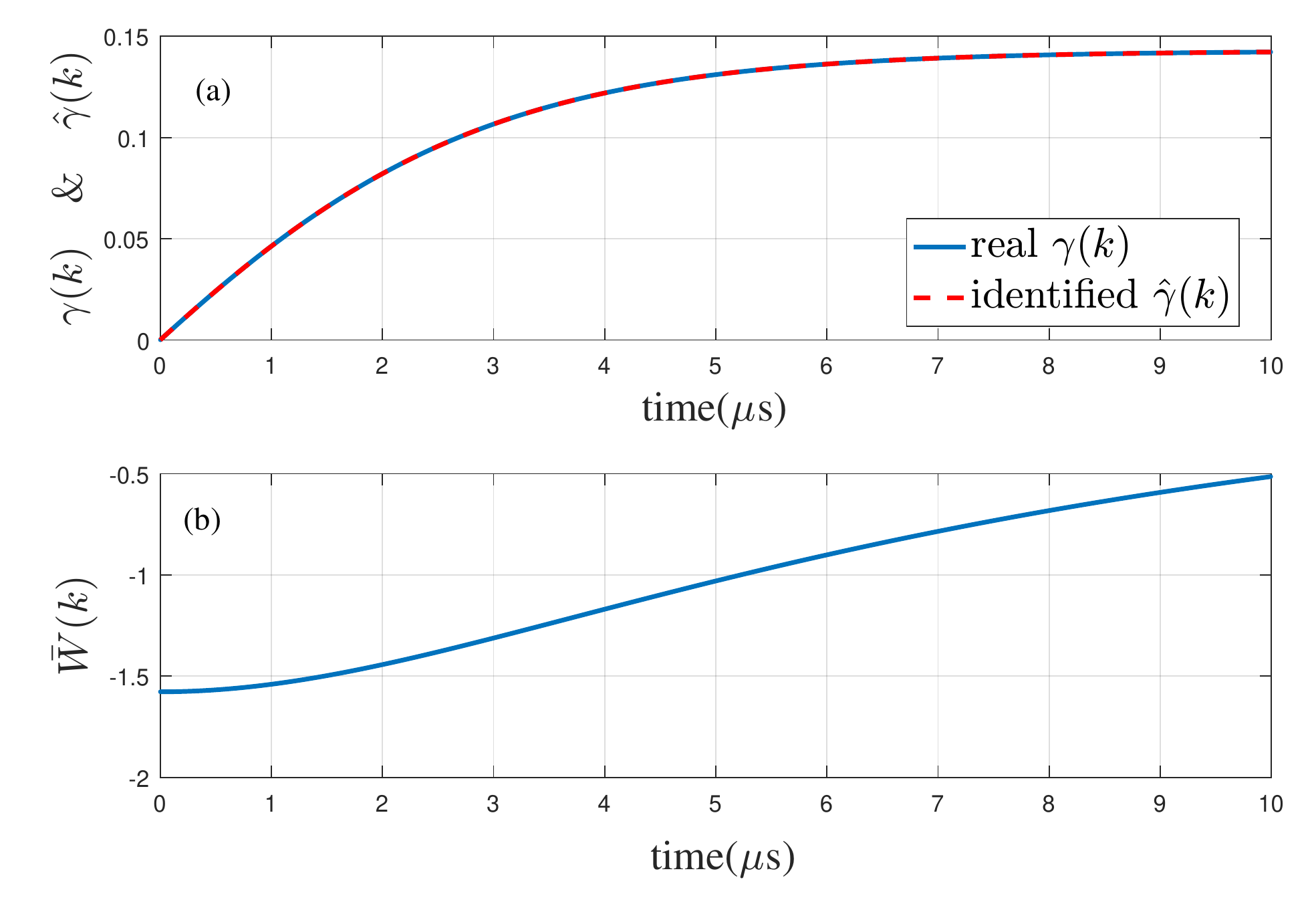}
	\caption{The identification results in the case of measuring ${\sigma}_z$ where our method can identify the real damping rate functions with nonzero $\bar{W}$.}	\label{sigmaz}
\end{figure}

\begin{figure}
	\centering	\includegraphics[width=2.5in]{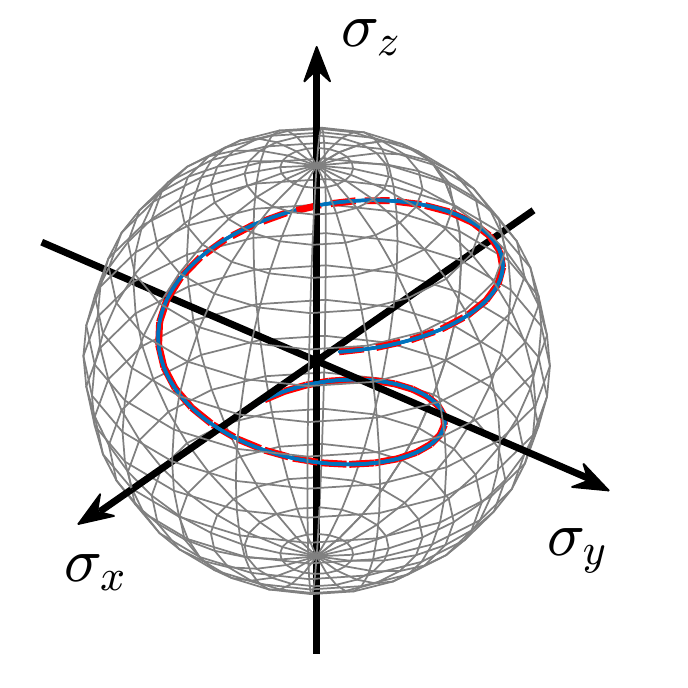}
	\caption{Evolution of the atom on the Bloch sphere where the calculated trajectory with the identified damping rate function (red dashed line) can perfectly match the real one (blue solid line).}
	\label{sigmaz_evolution}
\end{figure}

When we measure the observable ${\sigma}_z$; i.e., the output is $y(t)=\langle{\sigma}_z(t)\rangle$,
the identified damping rate function plotted as the red dashed line perfectly match the real one plotted as the blue solid line as shown in Fig. \ref{sigmaz}. There is no singularity since $\bar{W}$ is not zero during the identification process. Hence, the evolution of the atom induced by the identified damping rate function can match that with the real one on the Bloch sphere as shown in Fig.~\ref{sigmaz_evolution}. 
%
%

\begin{figure}
 	\centering
 		\includegraphics[width=2.8in]{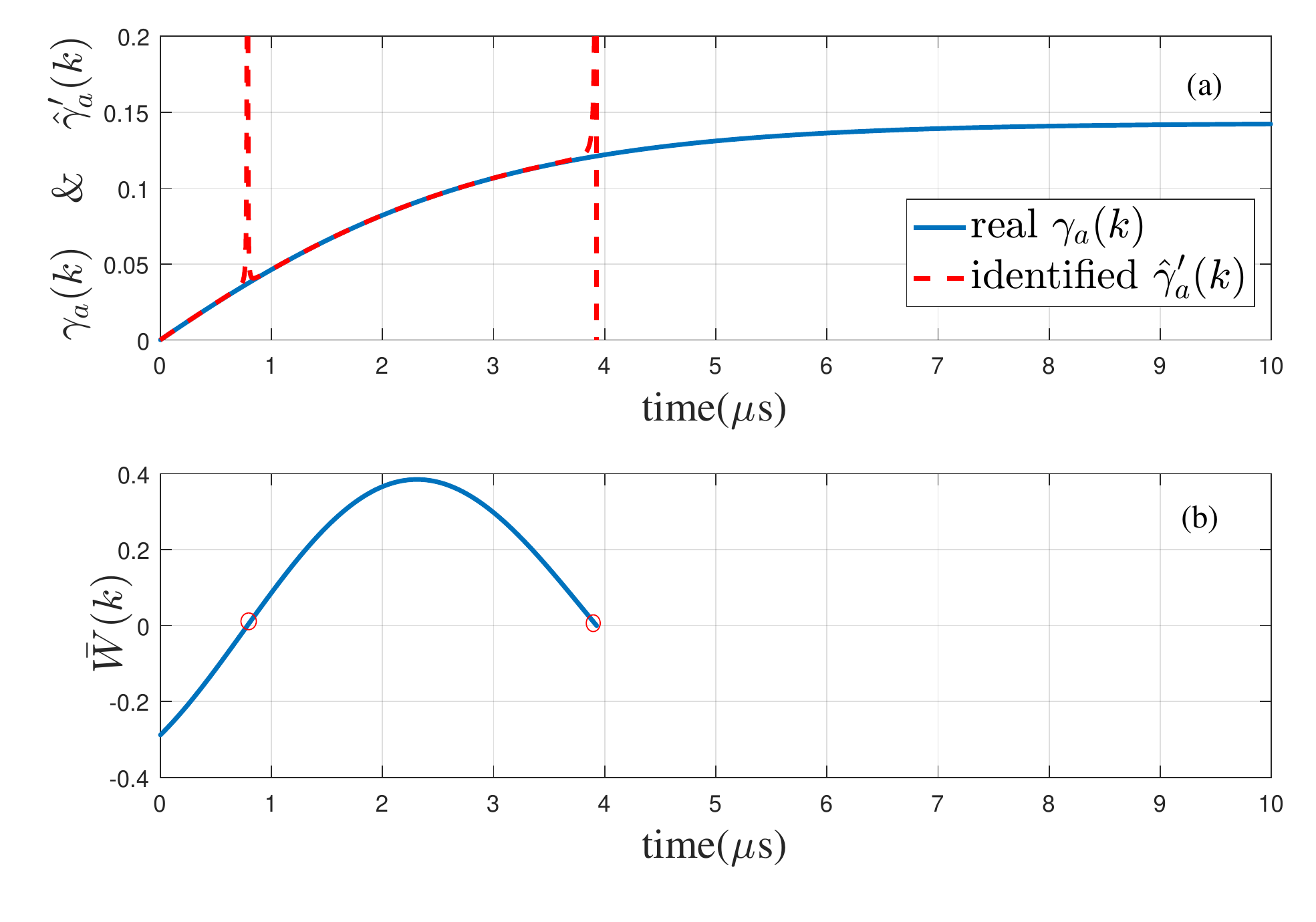}
 		\caption{The divergent identified damping rate function resulting from the singularity of $\bar{W}(k)$ when measuring $\sigma_x$.} 		\label{sigmax_singular}
  \end{figure}

\begin{figure}
	\centering
		\includegraphics[width=2.8in]{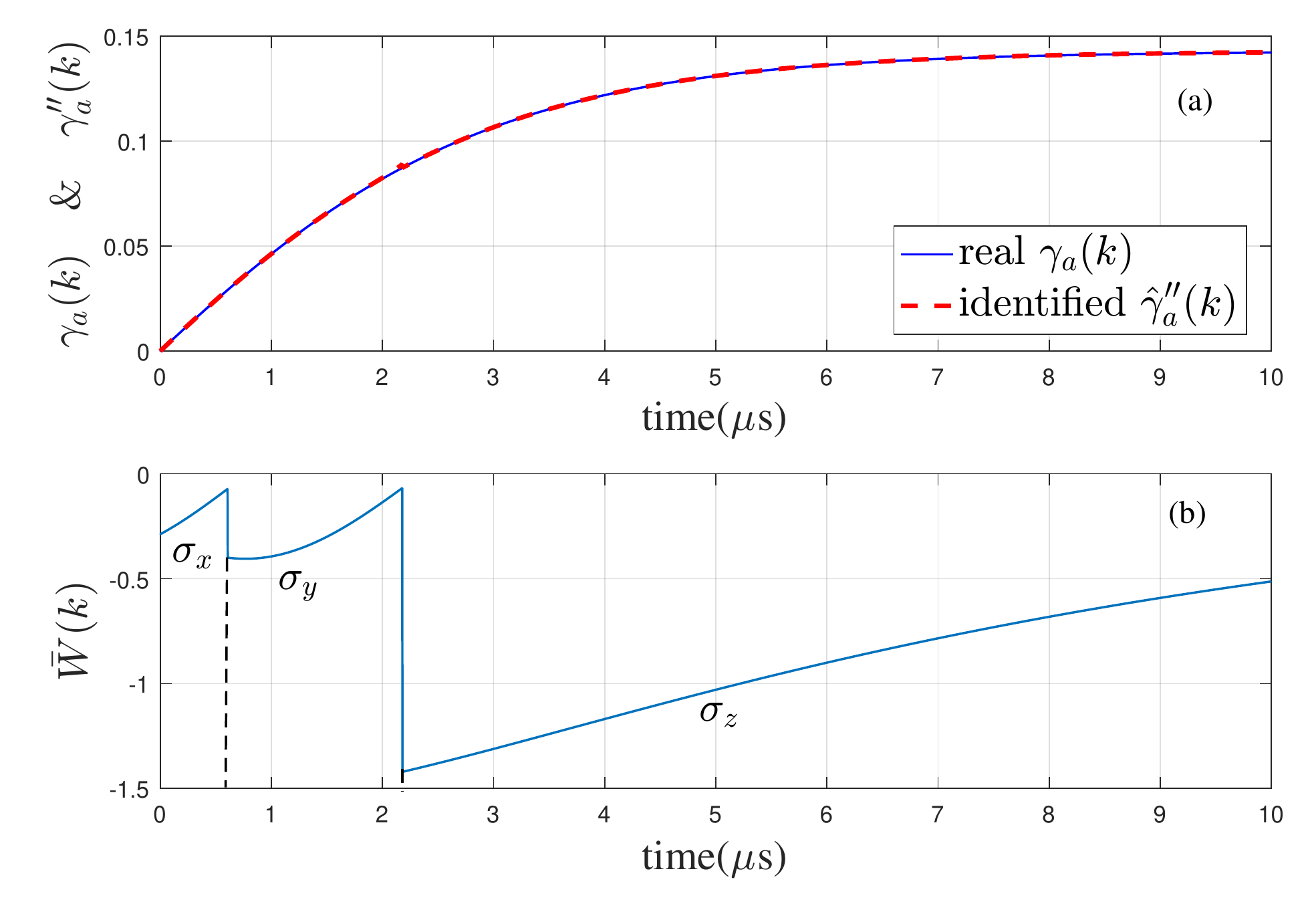}
		\caption{The identification results obtained by initially measuring ${\sigma}_x$£¬ where the singularity of $\bar{W}$ is avoided by changing the observables.}
		\label{sigma_xy}
	\end{figure}

\begin{figure}
		\centering
		\includegraphics[width=2.5in]{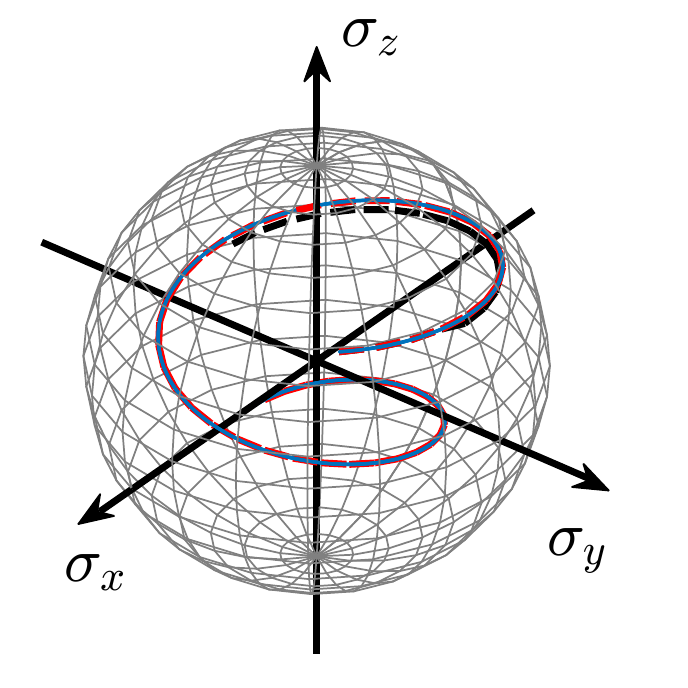}
		\caption{The evolution of the atom on the Bloch sphere with real damping rate functions (blue solid line) and the identified evolution plotted as the black dashed line and the red dashed line when measuring $\sigma_x$ only and multi-observables, respectively.}\label{sigmaxy_evolution}
\end{figure}

When we initially measure the observable ${\sigma_x}$; i.e., the output is $y(t)=\langle{\sigma}_x(t)\rangle$, 
$\bar{W}$ can be singular at some critical points such that the identified damping rate function $\hat{\gamma}'_a$ is divergent as shown in Fig. \ref{sigmax_singular}. 
To improve the identification result, we can change the observable when $\bar{W}$ is singular.
Here, we introduce two additional observables ${\sigma}_y$ and ${\sigma}_z$ and we will replace ${\sigma}_x$ by one of them to avoid the singularity.  With this improvement, we obtain a perfect identification result $\hat{\gamma}''_a$ as shown in Fig. \ref{sigma_xy}. It shows that by changing the observable we can avoid the case when $\bar{W}$ is singular. With the identification results $\hat{\gamma}'_a$ and $\hat{\gamma}''_a$, we also plot the evolution of the atom in Fig. \ref{sigmaxy_evolution} where the trajectory resulting from $\hat{\gamma}''_a$ (red dashed line) is comparable to the real one (blue solid line) but that calculated by using $\hat{\gamma}'_a$ (black dashed line) has a discrepancy from the real one.


\subsection{Non-Markovian environment identification for a three-spin chain system}
Next we consider a three-spin chain system where each spin is in an independent non-Markovian environment. Our task is to simultaneously identify the three damping rate functions in the time-convolution-less master equation
\begin{equation}\label{master2}
  {\dot{\rho}}(t)=-i[{H}_{sp},{\rho}(t)]+\sum_{i=1}^{3}\gamma_i(t)\mathcal{L}_{\gamma_i}{\rho}(t).
\end{equation}
The Hamiltonian of the three-spin chain system is written as
\begin{eqnarray}\label{Hami}
  {H}_{sp}&=&\frac{1}{2}(\omega_1{\sigma}_z^1+\omega_2{\sigma}_z^2+\omega_3{\sigma}_z^3)+\frac{g_1}{2}({\sigma}_x^1{\sigma}_x^2+{\sigma}_y^1{\sigma}_y^2)\nonumber\\
  &&+\frac{g_2}{2}({\sigma}_x^2{\sigma}_x^3+{\sigma}_y^2{\sigma}_y^3),
\end{eqnarray}
where the spins are coupled in an $XY$ interaction fashion. The symbols $\omega_{1,2,3}$ are the splitting frequencies of the three spins. The symbols $g_1$ and $g_2$ are the coupling strengthes between the spin $1$ and $2$ and the spin $2$ and $3$, respectively. The superscripts of the Pauli matrices $\sigma$ label the spins.
The three dissipative channels for the spins are all in Lindblad form
\begin{eqnarray}\label{L0LD2}
\mathcal{L}_{i}{\rho}(t)&=&{\sigma}_-^i{\rho}(t){\sigma}_+^i-\frac{1}{2}{\sigma}_+^i{\sigma}_-^i{\rho}(t)-\frac{1}{2}{\rho}(t){\sigma}_+^i{\sigma}_-^i,\nonumber\\
&&~~~~~~~~~~~~~~~~~~~~~~~~~~~~~~~~~~~~~~i=1,2,3.
\end{eqnarray}

%
%
%
%
To simulate the dynamics of the system, we set the parameters for the spin system as $\omega_1=1{\rm GHz}$, $\omega_2=1.5{\rm GHz}$, $\omega_3=1.4{\rm GHz}$, $g_1=1{\rm GHz}$, and $g_2=4{\rm GHz}$. The initial density matrix is set to be $\rho(0)=\rho_1(0)\otimes\rho_2(0)\otimes\rho_3(0)$, where $\rho_i(0)=\frac{1}{2}(I+\frac{1}{\sqrt{3}}\sigma_x^i+\frac{1}{\sqrt{3}}\sigma_y^i+\frac{1}{\sqrt{3}}\sigma_z^i), i=1,2,3$.
We sample the observable thirty thousands times in a total time $10\mu s$.
We assume that the first and third time-varying damping rate functions $\gamma_1(t)$ and $\gamma_3(t)$ are of an identical form as
\begin{equation}\label{gamma1}
  \gamma_1(t)=\gamma_3(t)=\frac{2\gamma_{0l}\lambda_l\sinh(d_lt/2)}{d_l\cosh(d_lt/2)+\lambda_l\sinh(d_lt/2)},~l=1,3,
\end{equation}
with $\lambda_1=0.1{\rm GHz}, \gamma_{01}=0.5{\rm GHz}, d_1=0.6{\rm GHz}$ and $\lambda_3=0.5{\rm GHz}, \gamma_{03}=0.5{\rm GHz}, d_3=0.5{\rm GHz}$, respectively. The second damping rate function is in a different form as
\begin{eqnarray}\label{gamma2}
  \gamma_2(t)&=&\frac{\gamma_{02}\lambda_2^2}{\lambda_2^2+d_2^2}[1-e^{-\lambda_2 t}(\cos(d_2t)-\frac{d_2}{\lambda_2}\sin(d_2t))]\nonumber\\
              &&+\frac{\gamma_{02}^2\lambda_2^5e^{-\lambda_2 t}}{2(\lambda_2^2+d_2^2)^3}\{[1-3(\frac{d_2}{\lambda_2})^2](e^{\lambda_2 t}-e^{\lambda_2 t}\cos(2d_2t))\nonumber\\
              &&-2[1-(\frac{d_2}{\lambda_2})^4]\lambda_2 t\cos(d_2t)+4[1+(\frac{d_2}{\lambda_2})^2]d_2t\sin(d_2t)\nonumber\\
              &&+\frac{d_2}{\lambda_2}[3-(\frac{d_2}{t})^2]e^{-\lambda_2 t}\sin(2d_2t)\}
\end{eqnarray}
with $\gamma_{02}=0.3{\rm GHz}, \lambda_2=1{\rm GHz}, d_2=2.4{\rm GHz}$. These forms of damping rate functions can be found in~\cite{breuer2002theory}.



\begin{figure}
		\centering
		\includegraphics[width=3.2in]{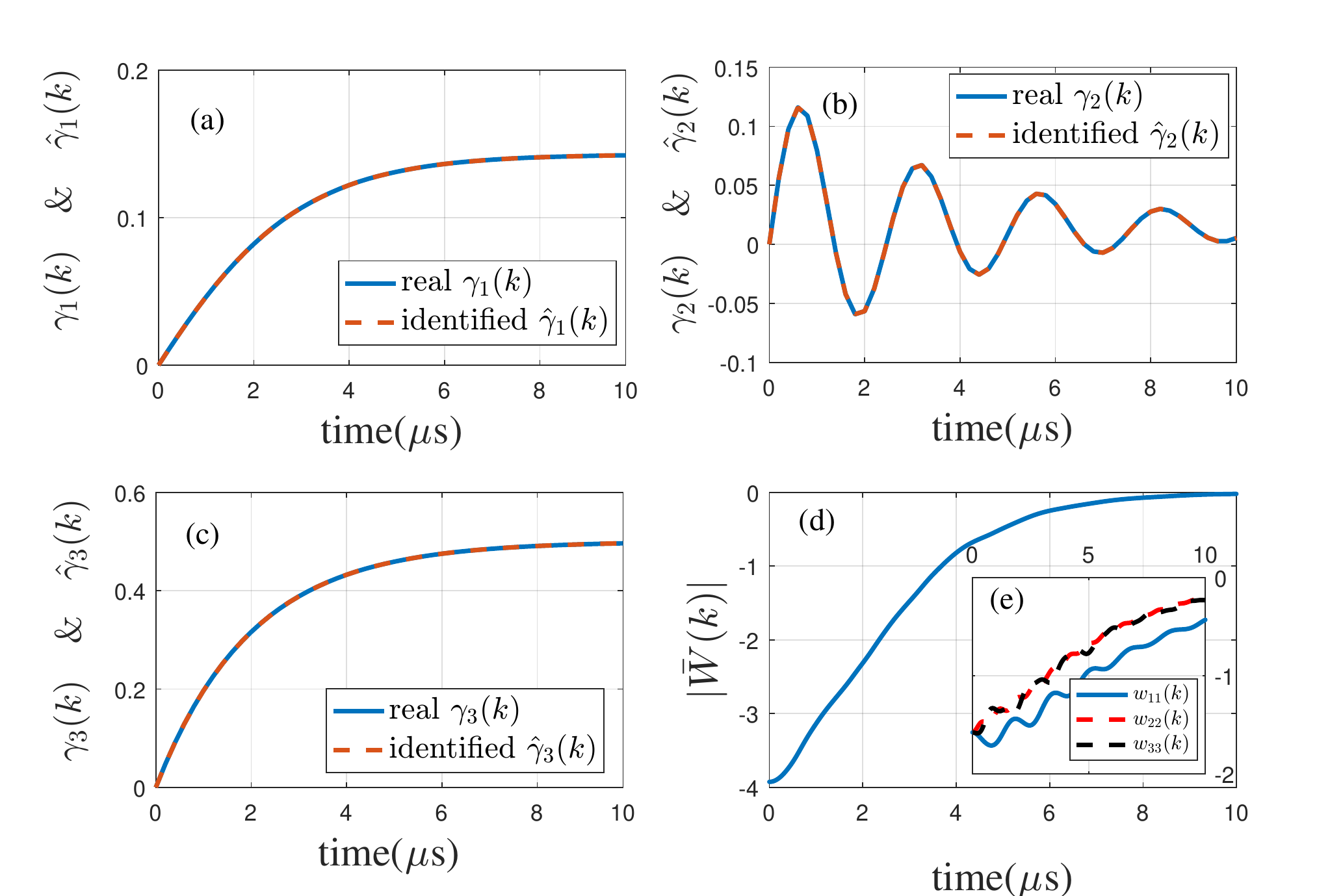}
		\caption{Identification results for the three damping rate functions where the real and identified damping rate functions are plotted as blue solid lines and red dashed lines, respectively, and the determinant and eigenvalues of $\bar{W}$ when measuring $\sigma^1_z, \sigma_z^2, \sigma_z^3$.}
		\label{r3}
\end{figure}

\begin{figure}
	\centering
	\includegraphics[width=3in]{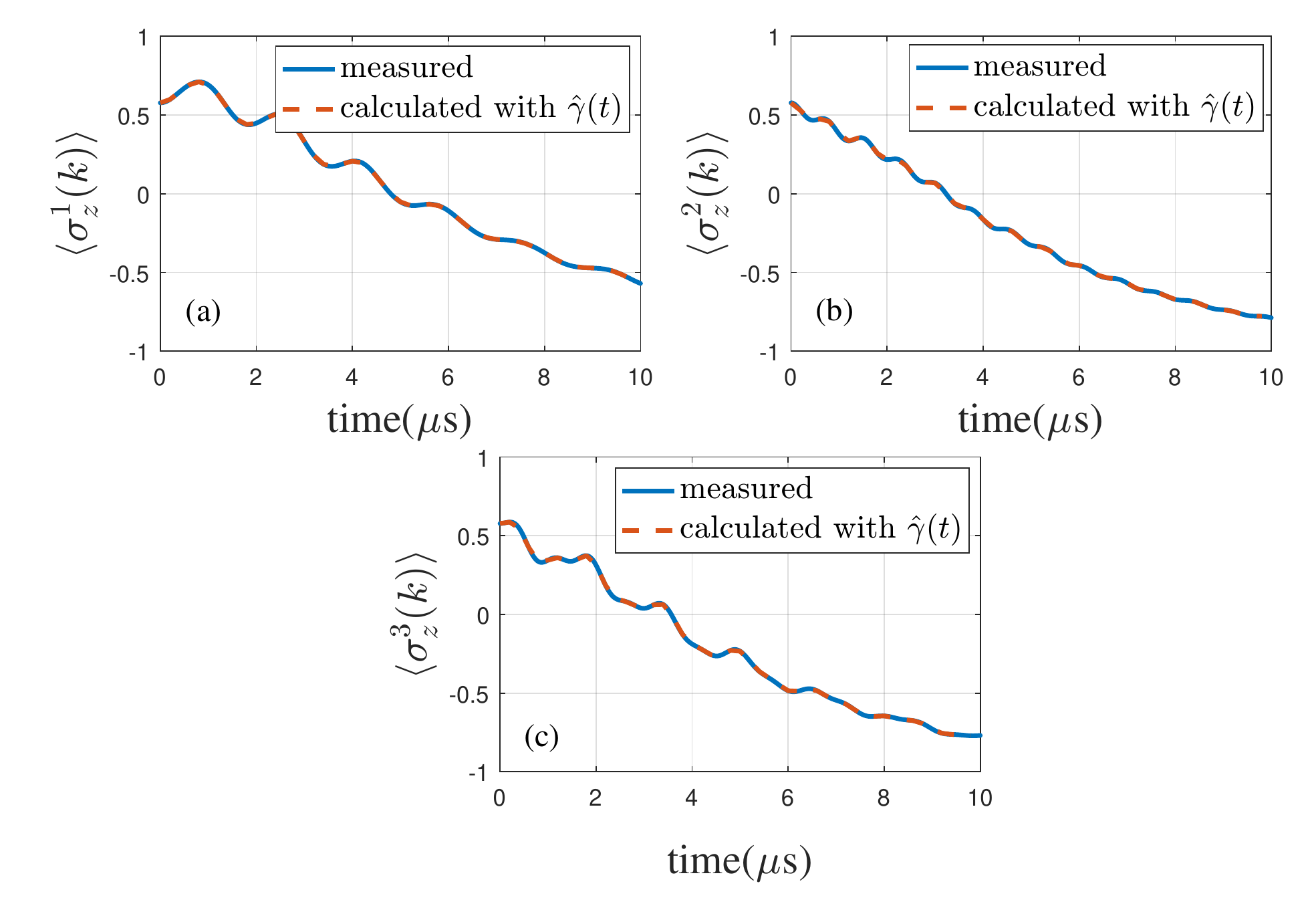}
	\caption{Comparison between the measurements (blue solid line) and the calculated evolutions of $\langle\sigma_z^1\rangle, \langle\sigma_z^2\rangle$ and $\langle\sigma_z^3\rangle$ (red dashed line) with the three identified damping rate functions.}
	\label{evolutionz}
\end{figure}

For this spin system, when we measure the observable $\sigma_z$ for each spin; i.e., the outputs of the system are written as $y_1(t)=\langle{\sigma}_z^1(t)\rangle$, $y_2(t)=\langle{\sigma}_z^2(t)\rangle$, and $y_3(t)=\langle{\sigma}_z^3(t)\rangle$, our algorithm can identify the three damping rate functions simultaneously as shown in Fig.\ref{r3}(a)-(c). This is because that the determinant of $\bar{W}$ is not zero due to the non-zero eigenvalues $w_{11}, w_{22}, w_{33}$ of $\bar{W}$ as shown in Fig. \ref{r3}(d) and (e). With these identified damping rate functions, the calculated evolutions of the three observables match the measurement results shown in Fig. \ref{evolutionz}.

\begin{figure}
	\centering
		\includegraphics[width=3.2in]{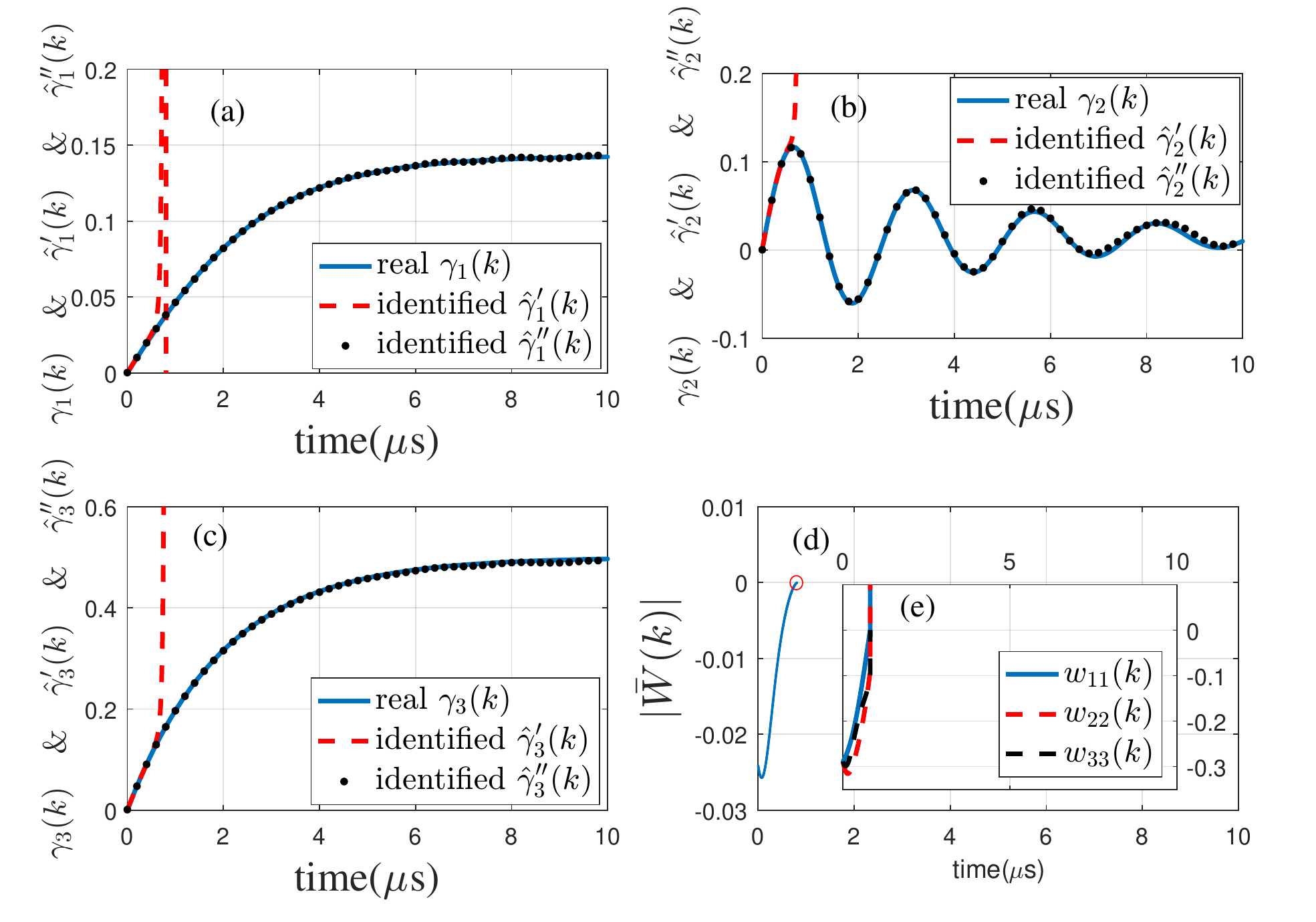}
		\caption{Comparison between three identified damping rate functions and the real ones (a)-(c) with the evolution of the determinant and eigenvalues of $\bar{W}(k)$ when measuring ${\sigma}_x^1, {\sigma}_x^2, {\sigma}_x^3$ (d) and (e).}		\label{gammaSingu}
\end{figure}

\begin{figure}
\centering
		\includegraphics[width=3.2in]{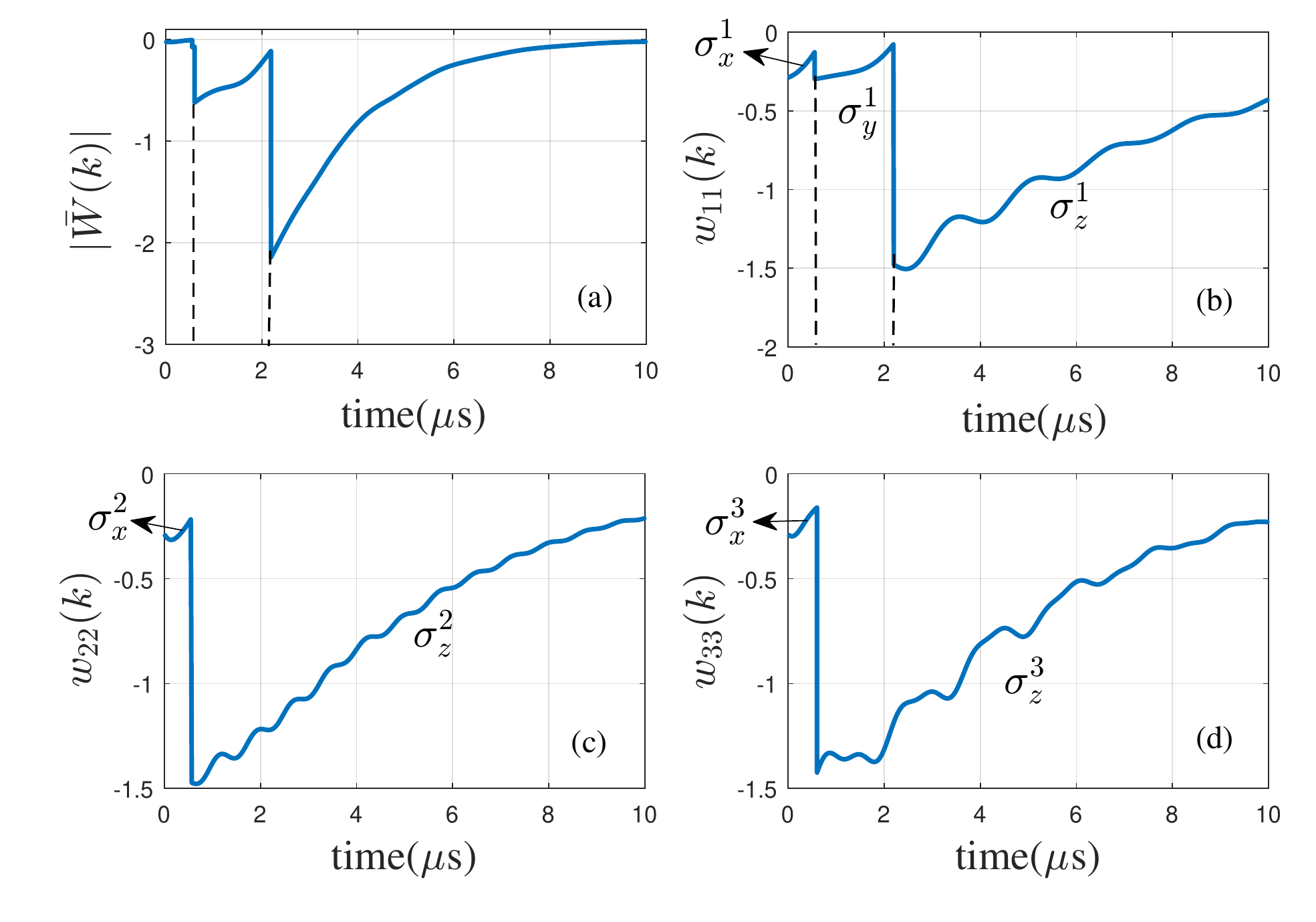}
		\caption{Variation of the determinant and the eigenvalues of $\bar{W}$ under the observable substitution strategy.}
		\label{Wx}
\end{figure}

\begin{figure}
	\centering
	\includegraphics[width=3in]{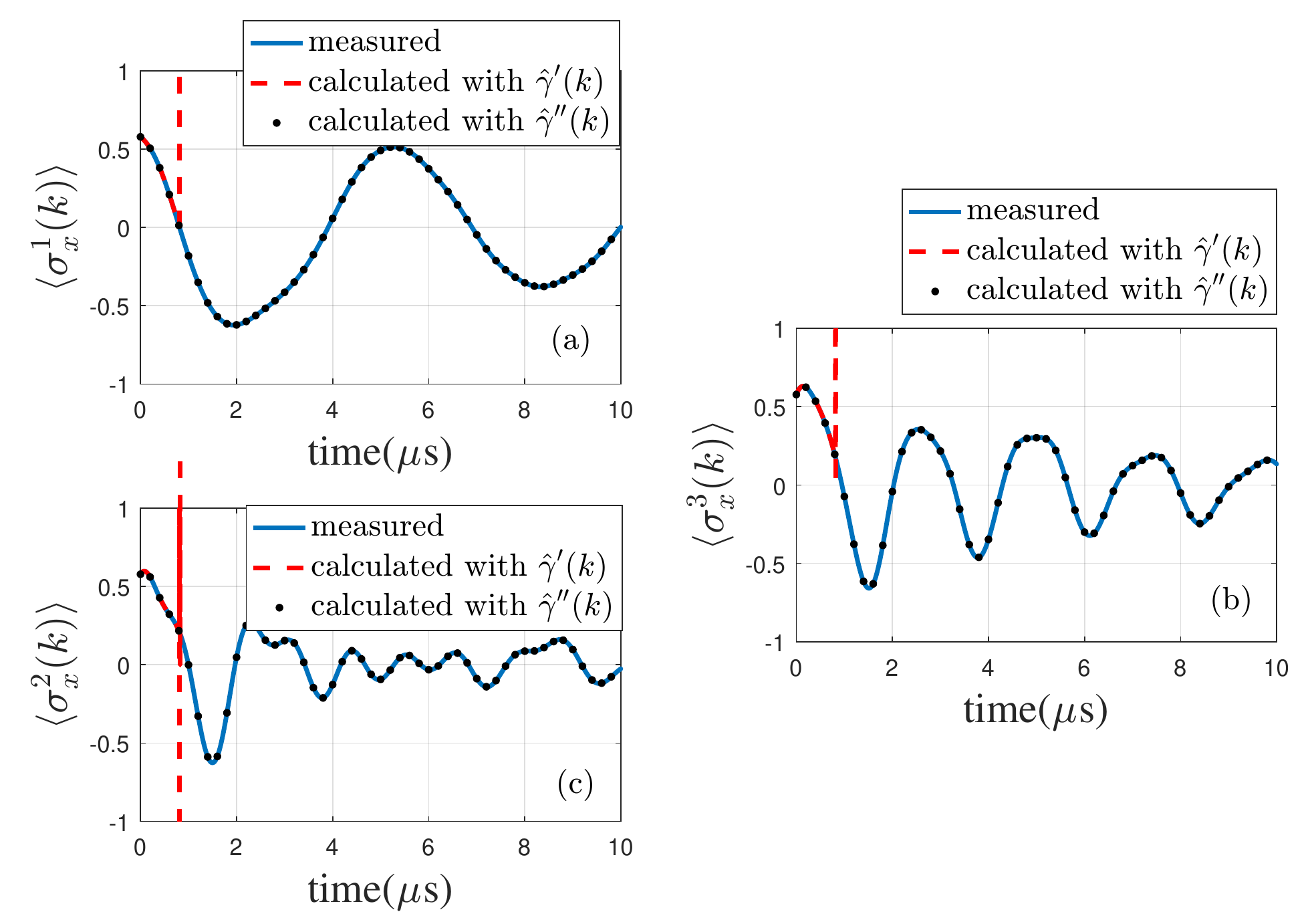}
	\caption{Evolutions of the time traces of the observables $\langle{\sigma}_x^1\rangle, \langle{\sigma}_x^2\rangle$ and $\langle{\sigma}_x^3\rangle$.}
	\label{evolutionx}
\end{figure}

In some circumstances, we can choose another set of observables, say ${\sigma}_x^1, {\sigma}_x^2, {\sigma}_x^3$. Using the time traces of these observables to identify the damping rate functions, we obtain the results plotted as red dashed lines in Fig.\ref{gammaSingu}, which deviate from the real ones plotted as blue solid lines. This is because the determinant of $\bar{W}$ becomes zero at a critical point. These identification results can be improved if we substitute the observable resulting in the singularity for a new one at these points. For example, we substitute the observable $\sigma_x^2$ for $\sigma_z^2$ as shown in Fig.\ref{Wx}(c). In this way, we obtain that the identification results plotted as black dots in Fig.\ref{gammaSingu} are consistent with the real damping rate functions plotted as blue solid lines. With these identification results, we calculate the evolution of the observables plotted as black dots in Fig.\ref{evolutionx}, which can match the measurements.





In summary, our algorithm can simultaneously identify multiple damping rate functions for dissipative channels in the TCL master equation. The cost of our algorithm is that we require sufficient measurements for the observables to avoid singularities. However, by using these measurement results, our algorithm saves computational time significantly compared to the gradient algorithm~\cite{xue2019identification}. This is because our algorithm solves the TCL master equation only once as well as reconstructing the damping rate functions interval by interval. In contrast, the number of times that the TCL master equation must be solved in the gradient algorithm is several thousands in order to ensure the solution converges to an optimal one.

\section{Conclusions}\label{VI}
We have presented an inverse-system method to identify the damping rate functions in a class of time-convolution-less master equations for non-Markovian quantum systems. This method can identify multiple damping rate functions simultaneously with sufficient measurements. A necessary condition for the identifiability of the damping rate functions is also given, with which we have designed a numerical algorithm. In our algorithm, we only calculate the evolution of the non-Markovian system once, such that the computational times are reduced significantly compared to gradient algorithms~\cite{xue2019identification}. Two examples of a non-Markovian single atom and a non-Markovian three-spin chain are given to show the effectiveness of our method.
\section*{Acknowledgment}
This work was supported in part by National Natural Science Foundation of China under Grants 61873162 and 61473199, in part by the Shanghai Pujiang Program under Grant 18PJ1405500, in part by the Open Research Project of the State Key Laboratory of Industrial Control Technology, Zhejiang University, China (No.ICT1900304), in part by the Suzhou Key Industry Technology Innovation
Project SYG201808, and in part by the Key Laboratory of
System Control and Information Processing in Ministry of Education of China Scip201804, and the Australian Research Council under grant DP180101805.
\appendix

\bibliographystyle{IEEEtran}
\bibliography{NMInverseIdentification}

\end{document}